\documentclass[journal, letter, 9pt]{IEEEtran}
\usepackage{cite}
\usepackage[Gray,amssymb]{SIunits}
\usepackage{color}
\usepackage{acronym}
\usepackage{esvect}
\usepackage[hidelinks]{hyperref}
\usepackage{amsfonts,amsmath,amssymb}
\usepackage{textcomp}
\usepackage{dsfont}
\usepackage{nicefrac}
\usepackage{lipsum}
\ifCLASSINFOpdf
  \usepackage[pdftex]{graphicx}
  \DeclareGraphicsExtensions{.pdf,.jpeg,.png}
\else
  \usepackage[dvips]{graphicx}
  \DeclareGraphicsExtensions{.eps}
\fi

\usepackage[hidelinks]{hyperref}

\usepackage{tikz}

\newcommand{\red}[1]{\textcolor{red}{#1}}

\newcommand{\T}{\mathrm{T}}

\DeclareGraphicsExtensions{.pdf,.jpeg,.png,.jpg}
\graphicspath{{./figures/}{../figures/}}

\acrodef{pfs}[{pfs}]{power flow solution}
\acrodef{pcc}[{pcc}]{point of common coupling}
\acrodef{poc}[{poc}]{point of connection}
\acrodef{opf}[{opf}]{optimal power flow}
\acrodef{hb}[{hb}]{harmonic balance}
\acrodef{smps}[{smps}]{switched mode power supply}
\acrodef{mvf}[{mvf}]{multi-variate formulation}
\acrodefplural{smps}[{smps}]{switched mode power supplies}
\acrodef{pll}[PLL]{phase locked loop}
\acrodef{dae}[DAE]{differential algebraic equation}
\acrodef{ode}[ODE]{ordinary differential equation}
\acrodef{sde}[SDE]{stochastic differential equation}
\acrodef{bvp}[BVP]{boundary value problem}
\acrodef{lte}[{lte}]{local truncation error}
\acrodef{lcm}[{lcm}]{least common multiple}
\acrodef{rmse}[{rmse}]{root mean squared error}
\acrodef{vc}[{vc}]{virtual connector}
\acrodef{ps}[{pss}]{power electromechanical sub-system}
\acrodef{ess}[{ess}]{electromagnetic sub-system}
\acrodef{mna}[{mna}]{modified nodal analysis}
\acrodef{psm}[PSM]{power system model}
\acrodef{emt}[{emt}]{electromagnetic transient}
\acrodef{ts}[{ts}]{transient stability}
\acrodef{shpf}[{shpf}]{shooting power flow}
\acrodef{pf}[{pf}]{power flow}
\acrodef{vsc}[{vsc}]{voltage source converter}
\acrodef{mppt}[{mppt}]{maximum power point tracker}
\acrodef{pic}[{pi}]{proportional-integral}
\acrodef{ibr}[{ibr}]{inverter-based resource}
\acrodef{sdae}[SDAE]{Stochastic Differential Algebraic Equation}
\acrodef{ou}[OU]{Ornstein-Uhlenbeck}
\acrodef{mc}[MC]{Measure of Closeness}

\newcommand{\bfb}[1]{\boldsymbol{\rm #1}}

\newcommand{\Dt}{\dot}

\newcommand{\PreserveBackslash}[1]{\let\temp=\\#1\let\\=\temp}
\newcommand{\bfg}[1]{\boldsymbol{#1}}
\newcommand{\bft}[1]{\tilde{\boldsymbol{#1}}}

\newcommand{\EM}{Euler-Maruyama}
\newcommand{\sm}{the synchronous machines, respectively}
\newcommand{\covz}{\bfb C}
\newcommand{\covy}{\bfb K}
\newcommand{\MC}{measure of closeness}

\IEEEoverridecommandlockouts

\newcommand\copyrighttext{%
  \footnotesize
  \centering\copyright~2022 IEEE. Personal use of this material is permitted. Permission from IEEE must be obtained for all other uses, in any current or future media, including reprinting/republishing this material for advertising or promotional purposes, creating new collective works, for resale or redistribution to servers or lists, or reuse of any copyrighted component of this work in other works.\\
  IEEE Trans. on Power Sys.
  DOI:\href{https://doi.org/10.1109/TPWRS.2022.3226076}{10.1109/TPWRS.2022.3226076}}
\newcommand\copyrightnotice{%
\begin{tikzpicture}[remember picture,overlay]
\node[anchor=north,yshift=0pt] at (current page.north)
{\setlength{\fboxrule}{0pt}\fbox{\parbox{\dimexpr\textwidth-\fboxsep-\fboxrule\relax}{\copyrighttext}}};
\end{tikzpicture}%
}

\begin{document}
\title{On the Calculation of the Variance of Algebraic Variables in
  Power System Dynamic Models with Stochastic Processes}

\author{%
  Muhammad Adeen,~\IEEEmembership{IEEE Student Member}, Federico
  Bizzarri,~\IEEEmembership{IEEE Senior Member}, \\ Davide del
  Giudice,~\IEEEmembership{IEEE Student Member}, Samuele
  Grillo,~\IEEEmembership{IEEE Senior Member}, \\ Daniele
  Linaro,~\IEEEmembership{IEEE Member}, Angelo
  Brambilla,~\IEEEmembership{IEEE Member}, Federico
  Milano,~\IEEEmembership{IEEE Fellow}\vspace{-7mm}%
  \thanks{F.~Bizzarri, D.~del Giudice, S.~Grillo, D.~Linaro, and
    A.~Brambilla are with Politecnico di Milano, DEIB, p.zza Leonardo
    da Vinci, n. 32, 20133 Milano, Italy.  (e-mails:
    \{federico.bizzarri, davide.delgiudice, samuele.grillo,
    daniele.linaro, angelo.brambilla\}@polimi.it).}%
  \thanks{F.~Bizzarri is also with with the Advanced Research Center
    on Electronic Systems for Information and Communication
    Technologies E.~De Castro (ARCES), University of Bologna, 41026
    Bologna, Italy.}%
  \thanks{M.~Adeen and F.~Milano are with School of Electrical and
    Electronic Engineering, University College Dublin, Belfield
    Campus, Dublin 4, Ireland.  (e-mails:
    muhammad.adeen@ucdconnect.ie, federico.milano@ucd.ie)}%
  \thanks{M.~Adeen and F.~Milano are by the European Commission,
    supported, under the project EdgeFLEX, grant no.~883710.
    F.~Milano is also supported the Sustainable Energy Authority of
    Ireland (SEAI) under the project FRESLIPS, grant no.~RDD/00681.}%
  \thanks{Italian MIUR project PRIN~{2017K4JZEE\_006} funded the work
    of S.~Grillo (partially) and D.~del Giudice (totally).}  }

\IEEEaftertitletext{\copyrightnotice\vspace{0.2\baselineskip}}
\maketitle

\begin{abstract}
  This letter presents a technique to calculate the variance of
  algebraic variables of power system models represented as a set of
  stochastic differential-algebraic equations.  The technique utilizes
  the solution of a Lyapunov equation and requires the calculation of
  the state matrix of the system.  The IEEE 14-bus system serves to
  demonstrate the accuracy of the proposed technique over a wide range
  of variances of stochastic processes.  The accuracy is evaluated by
  comparing the results with those obtained with Monte Carlo time
  domain simulations.  Finally, a case study based on a 1479-bus
  dynamic model of the all-island Irish transmission system shows the
  computational efficiency of the proposed approach compared to the
  Monte Carlo method.
\end{abstract}


\section{Introduction}
\label{sec:intro}

\IEEEPARstart{M}{odern} power systems are subjected to stochastic
processes due to the high penetration of non-synchronous generation
such as wind and photo-voltaic.  Stochastic load consumption is
another relevant source of noise, especially at the distribution
level.  It is important to study the impact of stochastic processes to
estimate the probability that physical limits such as voltage
insulation ratings of a substation, the thermal limits of the
lines/transformers, are violated in normal operation.  This letter
focuses precisely on this point and proposes an efficient technique
that, given the properties of the noise sources, evaluates the
variance of all state and algebraic variables of power system dynamic
models.

The dynamic performance of a power system with inclusion of stochastic
processes can be conveniently studied through \acp{sdae}
\cite{milano2013systematic}.  These \acp{sdae} are non-linear and can
have high dimensionality for large power systems.  The use of
numerical schemes for their integration is thus unavoidable.  The
stochastic terms require a significant extra computational burden to
solve the integration \cite{RODEBOOK}.  Moreover, \acp{sdae} have to
be studied with a Monte Carlo method, i.e., several hundreds or even
thousands of times, to properly estimate the statistical properties of
the system variables, such their probability distribution and
variance.

A number of techniques are available in the literature that provide
the statistical properties in stationary conditions of the
\textit{state variables}, e.g.~\cite{vorobev2019deadbands} and
\cite{stoinit}.  These methods are based on the properties of the
Fokker-Planck equation and the solution of the Lyapunov equation.
This letter further elaborates on this approach and proposes the
utilization of a simple, yet effective linearized method to calculate
the statistical properties of the \textit{algebraic variables} of
power systems without the need to perform cumbersome time domain
simulations.

\section{Power System Model with Noise}
\label{sec:model}

The power system model considered in this work is described by the following set of
index-1 \acp{sdae}:
\begin{align}
  \Dt{\bfg x} &= \bfg f(\bfg x, \bfg y, \bfg \eta) \, , \label{eq:xeq} \\
  \bfg 0_{m,1} &= \bfg g(\bfg x, \bfg y, \bfg \eta) \, , \label{eq:yeq} \\
  \Dt{\bfg \eta} &= \bfg a(\bfg \eta) + \bfb B(\bfg \eta) \, \bfg \xi \, .\label{eq:neq}
\end{align}
Equations \eqref{eq:xeq}-\eqref{eq:yeq} describe the conventional
deterministic power system models such as transmission lines,
generators and controllers.  Vector
$\bfg f: \mathbb{R}^{n+m+p} \mapsto \mathbb{R}^n$ defines the
deterministic differential equations; vector
$\bfg g: \mathbb{R}^{n+m+p} \mapsto \mathbb{R}^m$ defines the
algebraic equations; $\bfg x\in \mathbb{R}^n$ is the vector of the
deterministic state variables and $\bfg y \in \mathbb{R}^m$ is the
vector of the algebraic variables.  Equation \eqref{eq:neq} defines
the behavior of the stochastic processes $\bfg \eta \in \mathbb{R}^p$,
where $\bfg \xi \in \mathbb{R}^q$ is the white noise vector, i.e., the
vector of the formal time derivatives of the Wiener processes.  Vector
$\bfg \eta$ represents the fluctuations of loads and renewable energy
sources such as wind and solar power plants.  Equation \eqref{eq:neq}
is composed of two terms: the drift
$\bfg a: \mathbb{R}^{p} \mapsto \mathbb{R}^p$ and the diffusion term
$\bfb B: \mathbb{R}^{p} \mapsto \mathbb{R}^p \times \mathbb{R}^{q}$.
The elements of matrix $\bfb B$ are, in general, nonlinear functions
of $\bfg \eta$ to account for distributions other than Gaussian
\cite{JONSDOTTIR2019368}.  However, for short-term analysis and/or
small fluctuations, one can assume normal distributions, which lead to
a constant diffusion matrix $\bfg B$ \cite{milano2013systematic}.

In the general case, $p \ne q$.  This feature allows modeling
correlated processes \cite{ADEEN9380527}.
If $p = q$ and $\bfb B$ is a diagonal matrix, then the processes
$\bfg \eta$ are fully uncorrelated.  In the following, without lack of
generality, the elements of $\bfg \eta$ are assumed to be uncorrelated
and with bounded variance.  The latter property has been observed in
the measurements of the noise sources of power systems, as discussed
in \cite{milano2013systematic} and references therein.  A widely
accepted model with bounded variance is the mean-reverted process,
which has a linear drift term of the form:
\begin{equation}
  \label{eq:mr}
  a_k(\eta_k) = \alpha_k (\mu_k - \eta_k) \, ,
  \qquad k = 1, 2, \dots, p\, ,
\end{equation}
where $\alpha_k$ is the mean reversion speed, which defines the
autocorrelation of the process, and $\mu_k$ is the average value of
the process.  If a mean-reverted drift \eqref{eq:mr} is coupled with a
constant diffusion, say $b_k$, one obtains a normally distributed
\ac{ou} process with variance:
\begin{equation}
  \label{eq:OUvar}
  {\rm var}(\eta_k) = \sigma^2_k = {b^2_k}/({2\alpha_k}) \, ,
  \qquad k = 1, 2, \dots, p\, ,
\end{equation}
where $\sigma_k$ is the standard deviation of the $k$-th process.
Note also that $b_k$ is the $k$-th diagonal element of the diffusion
matrix $\bfb B$.
%

\section{Calculation of the Variance of Algebraic Variables}
\label{sec:variance}

The starting point of the proposed technique to calculate the
variances of the algebraic variables $\bfg y$ is the set of \acp{sdae}
linearized at the equilibrium point
$(\bfg x_o, \bfg y_o, \bfg \eta_o)$ as per Method I described in
\cite{stoinit}, i.e., a point for which \eqref{eq:yeq} are satisfied
and $\dot {\bfg x} = \bfg 0_{n, 1}$ and
$\bfg a(\bfg \eta_o) = \bfg 0_{p,1}$.  The linearization of
\eqref{eq:xeq}-\eqref{eq:neq} gives:
\begin{equation}
  \label{eq:SDAElinear}
  \begin{aligned}
    \begin{bmatrix}
      \Dt{\bft x} \\
      \bfg 0_{m,1} \\
      \Dt{\bft \eta}
    \end{bmatrix}
    =
    \begin{bmatrix}
      \bfg f_{\bfg x} & \bfg f_{\bfg y} & \bfg f_{\bfg \eta} \\
      \bfg g_{\bfg x} & \bfg g_{\bfg y} & \bfg g_{\bfg \eta} \\
      \bfg 0_{p,n} & \bfg 0_{p,m} & \bfg a_{\bfg \eta} \\
    \end{bmatrix}
    \begin{bmatrix}
      \bft x \\
      \bft y \\
      \bft \eta
    \end{bmatrix}
    +
    \begin{bmatrix}
      \bfg 0_{n,q} \\
      \bfg 0_{m,q} \\
      \bfb B(\bfg \eta_o)
    \end{bmatrix} \bfg \xi \, ,
  \end{aligned}
\end{equation}
where $\bfg f_{\bfg x}$, $\bfg f_{\bfg y}$, $\bfg f_{\bfg \eta}$,
$\bfg g_{\bfg x}$, $\bfg g_{\bfg y}$, $\bfg g_{\bfg \eta}$,
$\bfg a_{\bfg \eta}$ are the Jacobian matrices of the system
calculated at $(\bfg x_o, \bfg y_o, \bfg \eta_o)$.   $\bft x$ and
$\bft \eta$ represent the deterministic and the stochastic states of
the linearized system.   Eliminating the algebraic variables from
\eqref{eq:SDAElinear} and defining $\bft z = [\bft x\red{^\T}, \bft \eta\red{^\T}]^T$ leads
to a set of linear \acp{sde}, as follows:
\begin{align}
  \nonumber
  \begin{bmatrix}
    \Dt{\bft x} \\
    \Dt{\bft \eta}
  \end{bmatrix}
  &=
    \begin{bmatrix}
      \bfg f_{\bfg x} - \bfg f_{\bfg y}\bfg g^{\bfg -1}_{\bfg y} \bfg g_{\bfg x} &
      \bfg f_{\bfg \eta} - \bfg f_{\bfg y}\bfg g^{\bfg -1}_{\bfg y} \bfg g_{\bfg \eta} \\
      \bfg 0_{p,n} & \bfg a_{\bfg \eta}\\
    \end{bmatrix}
  \begin{bmatrix}
    \bft x \\
    \bft \eta
  \end{bmatrix}
  +
  \begin{bmatrix}
    \bfg 0_{n,q} \\
    \bfb B(\bfg \eta_o)
  \end{bmatrix} \bfg \xi\\
  &=\bfb A_o \, \bft z + \bfb B_o \, \bfg \xi \, .
  \label{eq:SDAElinear1}
\end{align}

Based on the Fokker-Planck equation, the probability distribution
$\bfg \varpi(\bft z)$ of all state variables in stationary condition
satisfies \cite{vorobev2019deadbands}:
\begin{equation}
  \bfg \varpi(\bft z) = ({\rm det}\mid 2\pi \covz \mid )^{-1/2}\cdot
  {\rm exp}\bigg(-\frac{1}{2}\bft z^\T \covz^{\bfg -1} \bft z\bigg) \, ,
  \label{eq:pdf}
\end{equation}
where $\covz$ is the covariance matrix of the state variables in
\eqref{eq:SDAElinear1}.  Matrix $\covz$ is symmetric and satisfies the
Lyapunov equation:
\begin{equation}
  \bfb A_o \covz + \covz \bfb A_o^\T = - \bfb B_o \bfb B_o^\T \, ,
  \label{eq:lyapunov}
\end{equation}
which is a special case of the Riccati equation.
The diagonal elements of $\covz$ are the steady-state variances of the
components of the state variables $\bft z$.  In particular, if the
\ac{ou} processes $\bfg \eta$ are not correlated, the last $p$
diagonal elements of $\covz$ are given by \eqref{eq:OUvar} where $a_k$
and $b_k$ are the $k$-th diagonal elements of $\bfg a_{\bfg \eta}$ and
$\bfb B_o$, respectively, and $\sigma_k^2$ are the variances of the
$p$ stochastic processes $\bft \eta$.  The interested reader can refer
to \cite{stoinit} for a comprehensive discussion on the numerical
solution of \eqref{eq:lyapunov}.

From \eqref{eq:SDAElinear1}, we observe that $\bft x$ can be written
as a linear combination of the entries of $\bft z$.  Hence, also the
elements of $\bft x$ are Gaussian processes.  Furthermore, the
covariance matrix $\covy$ of the small-signal algebraic variables can
be written as \cite{provost1992}:
\begin{equation}
  \covy = \bfb G_o \, \covz \, \bfb G_o^\T \, ,
  \label{eq:covy}
\end{equation}
where
\begin{equation}
  \bfb G_o = - \bfg g^{-1}_{\bfg y}
  \begin{bmatrix}
    \bfg g_{\bfg x} & \bfg g_{\bfg \eta}
  \end{bmatrix} .
\end{equation}
The diagonal elements of $\covy$ are the sought variances of the
algebraic variables $\bft y$.

Note that if $p \ll n$, i.e., the number of noise sources is much
smaller than the number of state variables, the covariance matrices
$\covz$ and hence $\covy$ might not be full rank.  A zero element in
the $k$-th position of the diagonal of $\covz$ ($\covy$) indicates
that the associated $\tilde{x}_k$ ($\tilde{y}_k$) are not affected by
noise.  In this case, the vector of stochastic processes $\bft z$ is
said to be {degenerate} \cite{Jacod2000}.

\section{Case Studies}
\label{sec:casestudy}

This section illustrates the accuracy and numerical efficiency of the
method in Section \ref{sec:variance} to calculate the variances of
algebraic variables of the power system.  All results are compared to
the ones obtained through Monte Carlo (MC) time domain simulations.
The power systems utilized in this case study are the IEEE 14-bus
system, and the all-island Irish transmission system (AIITS).  The MC
simulations are performed exploiting parallelism on 2
Intel\textregistered{} Xeon\textregistered{} CPUs at 2.20GHz with
20-cores each, running a Linux OS that exploits core virtualization
(hypertrading).  This means that at most 80 realizations were solved
in parallel for each MC simulation.  Equation \eqref{eq:lyapunov} is
solved using the open-source library SLICOT \cite{SLICOT}, whereas
time domain simulations are carried out with Dome \cite{6672387}.  In
the remainder of this section, the systems of equations
\eqref{eq:lyapunov} and \eqref{eq:covy} are referred to as Lyapunov
Equation Method (LEM).

In both power systems, the sources of noise are modeled as OU
processes and included in the loads and, for the AIITS, also in the
wind speeds.  Load consumption is modeled as voltage dependent
incorporating stochastic processes:
\begin{equation}
  \label{eq:loadstc}
  \begin{aligned}
    p_{\rm L}(t) &= (p_{{\rm L}0} + \eta_p(t)) (v(t)/v_0)^{\gamma} \, , \\
    q_{\rm L}(t) &= (q_{{\rm L}0} + \eta_q(t)) (v(t)/v_0)^{\gamma} \, , \\
    v_{\rm W}(t) &= v_{{\rm W}0} + \eta_w(t)  \, ,
  \end{aligned}
\end{equation}
where $p_{{\rm L}0}$ and $q_{{\rm L}0}$ are the active and reactive
power consumption at time $t=0$; $v(t)$ is the magnitude of the bus
voltage at the load bus; $v_0$ is the voltage magnitude at the load
bus at the start of the simulation; $\gamma$ defines the load voltage
dependence; and $\eta_p$ and $\eta_q$ are OU stochastic processes with
drift and diffusion terms defined as in \eqref{eq:mr} and
\eqref{eq:OUvar}, respectively.  $\gamma = 2$ is utilized in all
simulations.

In \eqref{eq:loadstc} $v_{{\rm W}0}$ is the initial wind speed; and
$\eta_w$ is a stochastic process characterized by a Weibull
distribution.  This is obtained using the technique described in
\cite{windsde}, namely, using the same drift term defined as in
\eqref{eq:mr} and a nonlinear diffusion function $b_w(\eta_w)$, as
follows:
\begin{equation}
  \label{eq:weibull}
  \begin{aligned}
    a_w(\eta_w) &= \alpha_w ( \mu_w - \eta_w ) \\
    b_w(\eta_w) &= \sqrt{b_1(\eta_w)  b_2(\eta_w)} \, ,
  \end{aligned}
\end{equation}
where
\begin{equation*}
  \begin{aligned}
    \mu_w &= \lambda \Gamma \left( 1 + \kappa^{-1} \right ) \, , \\
    b_1(\eta_w) &= 2  \alpha_w \, \eta_w \, c_1 \frac{\lambda}{{\kappa}}
  \left( c_2(\eta_w) \right)^{-\kappa} \, , \\
    b_2(\eta_w) &= \kappa  \exp
  \left(c_2^{\kappa}(\eta_w) \right)  \Gamma
  \left( 1 + c_1, c_2^{\kappa}(\eta_w) \right)
                  - \Gamma \left( c_1\right) \, .
  \end{aligned}
\end{equation*}
In the above equations, $c_1 = 1/\kappa$ and
$c_2(\eta_w) = \eta_w/\lambda$; $\alpha_w$ is the autocorrelation
coefficient; $\kappa$ and $\lambda$ are the shape and scale
parameters, respectively, of the Weibull distribution; and
$\Gamma(\cdot)$ and $\Gamma(\cdot, \cdot)$ are the Gamma function and
the Incomplete Gamma functions, respectively.


The stochastic processes are modeled with the following parameters:
${\alpha_p = \alpha_q = \alpha_w = 0.01}\,\text{s}^{-1}$;
${\sigma(\eta_p) = 0.05\,p_{\rm L0}}$;
${\sigma(\eta_q) = 0.05\,q_{\rm L0}}$ and
${\sigma(\eta_w) = 0.05\,v_{\rm W0}}$.  The integration of the
deterministic part of \acp{sdae} is performed by the implicit
trapezoidal method with the ${\Delta t = 0.01\,\text{s}}$ time step.
The OU processes are integrated using the \EM{} method with
${h = 0.01\,\text{s}}$ step size.

\subsection{IEEE 14-Bus System}

The IEEE 14-bus system contains 14 buses with 11 loads, 20
lines/transformers, and 5 synchronous machines.  The synchronous
generators are described by a sixth-order model, and are equipped with
turbine governors and IEEE Type-I automatic voltage regulators.  An
automatic generation control is also included in the model.  The IEEE
14-bus system is modified by including a wind power plant connected to
bus 5.  All device models and data can be found in
\cite{milano2010power}.

The IEEE 14-bus system is first simulated using the MC approach.  MC
simulations require the selection of two parameters, namely, the final
simulation time $t_f$ and the number $N$ of realizations of
$\bfg \eta$.  These parameters decide how the stationary conditions
are reached and, hence, directly impact the accuracy of the
statistical properties, such as the standard deviation, of the
variables.

\begin{figure}[t]
  \centering
  \includegraphics[width=0.9\linewidth]{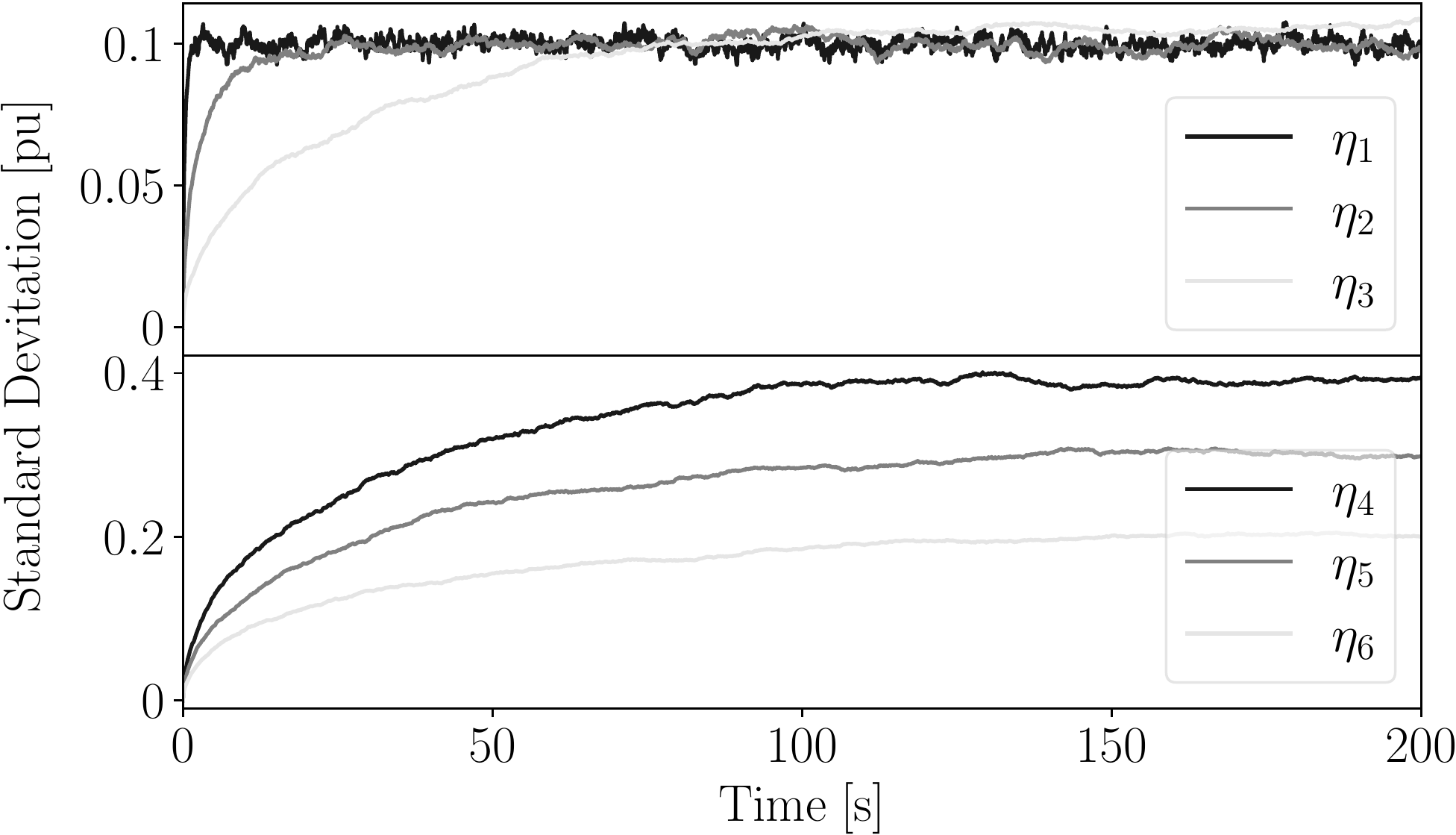}
  \caption{Standard deviation of $\eta$ as a function of time, where
    the parameters are defined such that
    $\sigma(\eta_1) = \sigma(\eta_2) = \sigma(\eta_3) = 0.1$ with
    $\alpha_1 = 1$ s$^{-1}$; $\alpha_2 = 0.1$ s$^{-1}$; and
    $\alpha_3 = 0.01$ s$^{-1}$, and $\sigma(\eta_4) = 0.4$;
    $\sigma(\eta_5) = 0.3$; $\sigma(\eta_6) = 0.2$ with
    $\alpha_4 = \alpha_5 = \alpha_6 = 0.01$ s$^{-1}$.}
  \label{fig:std}
  \vspace{-5mm}
\end{figure}

The choice of the value of $t_f$ depends principally on the parameters
$\alpha_i$ of the underlying processes and can be heuristically
estimated as $t_f = 2 / \min\{\alpha_i\}$.  As a proof of concept,
Fig.~\ref{fig:std} shows the time evaluation of $\sigma_{\eta}$ for
different values of $\alpha$.  Figure \ref{fig:std} shows that the
stationarity of $\sigma_{\eta}$ strongly depends on $\alpha$ (top
panel) and is independent from $\sigma_{\eta}$ at stationary
conditions (bottom panel).  More details on the evolution of the
standard deviation of the power system variables can be found in
\cite{9637935}.  In the simulations carried out for this case study,
the smallest $\alpha_i$ are of the order of $10^{-2}$ s and
Fig.~\ref{fig:std} confirms that a simulation time of $200$ s is
adequate to allow for all stochastic processes to reach stationarity.

\begin{figure}[b]
  \centering
  \vspace{-5mm}
  \includegraphics[width=0.99\linewidth]{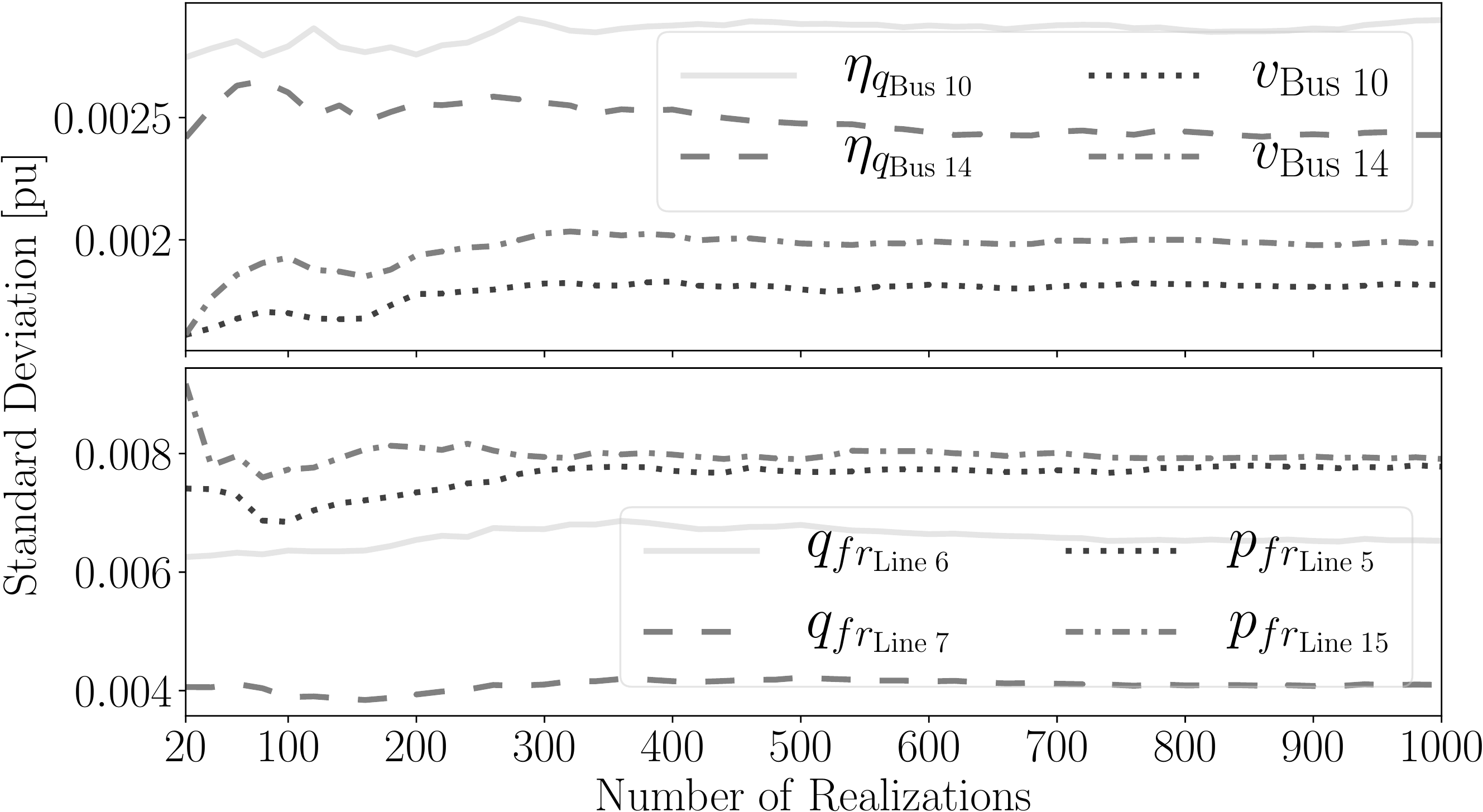}
  \caption{Standard deviation of variables as a function of the number
    of realizations of $\bfg \eta$.}
  \label{fig:variance}
\end{figure}

The second parameter to choose is the number of realizations $N$ for
which the estimated value of $\sigma_{\eta}$ is reliable enough to be
independent of the specific realizations of the stochastic processes.
The best value of $N$ is determined by calculating the average
$\sigma_{\eta}$ of each process at $t=200$ s as a function of $N$, as
shown in Fig.~\ref{fig:variance}.  This figure shows that as $N$
increases, the standard deviation of $\bfg \eta$ and of other system
variables converges towards a constant value.  Based on these results,
$N = 1000$ appears sufficient to obtain accurate stationary conditions
with the MC approach.

Next, we compare the values of standard deviation of the power system
variables for the IEEE 14-bus system obtained with MC with those
obtained through LEM.  With this aim, we define a \textit{measure of
  closeness}, $\epsilon_{\sigma}$, as follows:
\begin{equation}
  \label{eq:MC}
  \rm {\epsilon_{\sigma} \; (\%)} = \frac{\sigma_{\rm MC}-\sigma_{\rm LEM}}
  {\sigma_{\rm MC}}  100 \, .
\end{equation}
where $\sigma_{\rm MC}$, and $\sigma_{\rm LEM}$ are the standard
deviations of the variables obtained through MC and LEM, respectively.

\begin{figure}[t]
  \centering
  \includegraphics[width=0.9\linewidth]{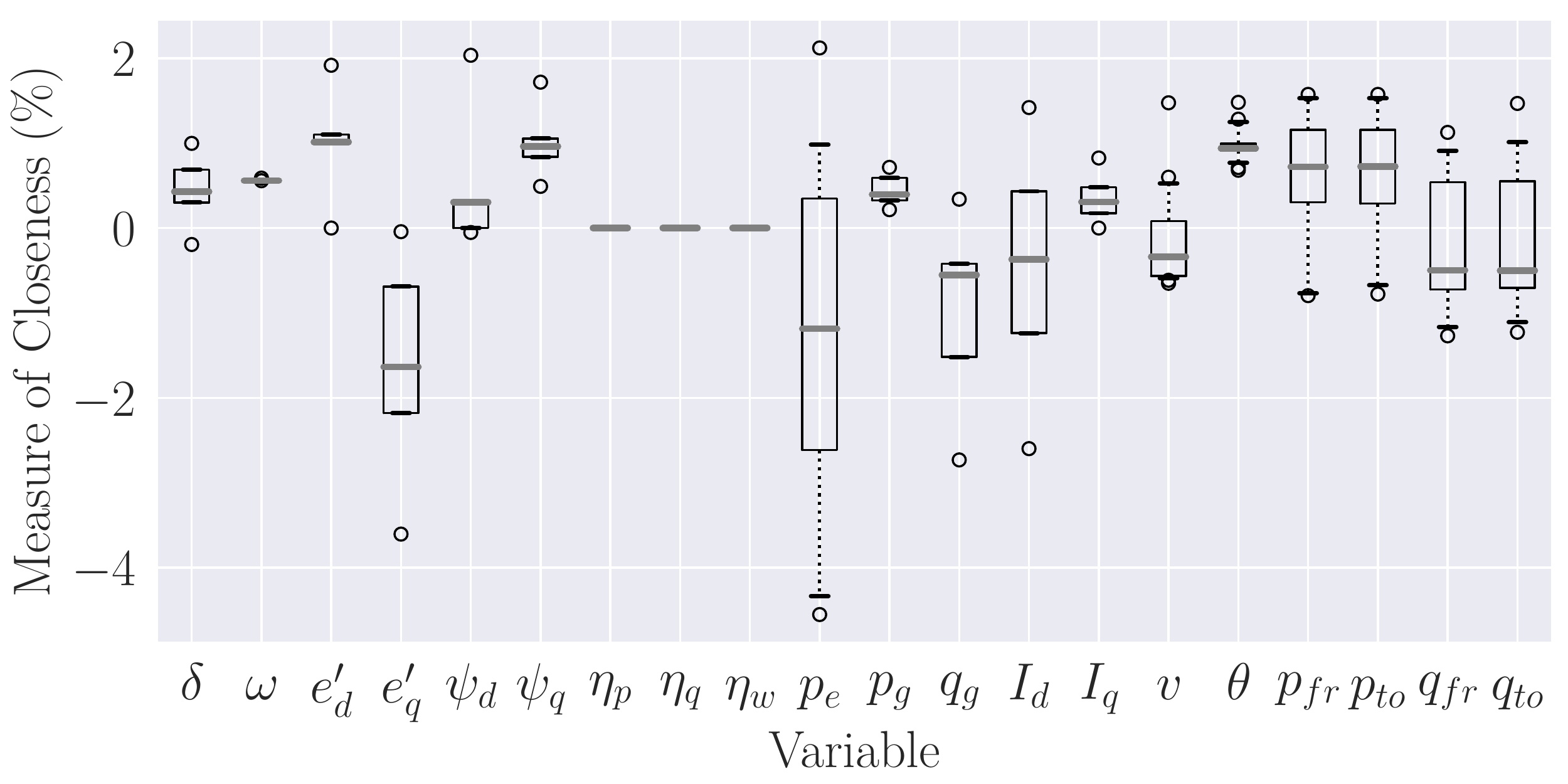}
  \caption{Box plot of the \MC{} $\epsilon_{\sigma}$ for the IEEE
    14-bus system.}
  \label{fig:ieee14_box}
  \vspace{-5mm}
\end{figure}

Figure \ref{fig:ieee14_box} shows the box plot of the values of
$\epsilon_{\sigma}$ obtained in the case of the IEEE 14-bus system
through the MC and LEM for the following variables: $\delta$ and
$\omega$ are the rotor angle and speed of \sm{}; $e'_d$ and $e'_q$
($\psi_d$ and $\psi_q$) are the d- and q-axis internal emfs (fluxes)
of \sm{}; $p_g$ and $q_g$ are the active and reactive power injections
of \sm{}; $p_e$ is the active power output of the wind power plants;
$I_d$ and $I_q$ are the d- and q-axis currents of \sm{}; $\theta$ and
$v$ are the bus voltage magnitude and angle, respectively; $p_{fr}$
and $q_{fr}$ ($p_{to}$ and $q_{to}$) are the active and reactive power
injections at the sending-end (receiving-end) bus, respectively.  In
the figure, the thick horizontal grey lines show the median of the
data, the top and bottom notches contain $5 \%$ to $95 \%$ percentile
of the data, and the black circles show the outliers.  Results
indicate that LEM yields $\sigma_{\rm LEM}$ that are very close to
$\sigma_{\rm MC}$.

Note that, for all the stochastic processes $\eta_p$, $\eta_q$ and
$\eta_w$, LEM yields the exact values of $\sigma_{\eta}$. This happens
regardless of the process being modeled through either a linear or a
non-linear diffusion term.  Note also that, to test the accuracy of
LEM against the nonlinearity of the \acp{sdae}, we have considered
$\sigma(\eta_p) = \sigma(\eta_q)$ ranging from $1 \%$ to $10 \%$ of
the initial load consumption.  The variations in the values of
$\epsilon_{\sigma}$ for all the variables were found to be in the same
range as in Fig.~\ref{fig:ieee14_box}.  It is fair to conclude, thus,
that LEM provides very accurate results for a wide range of standard
deviation of the stochastic process.

\subsection{All-Island Irish Transmission System}

This section demonstrates the robustness and light computational
burden of the LEM when applied to a real-world complex systems.  The
model of the AIITS considered in this section consists of 1479 buses,
1851 transmission lines or transformers, 245 loads, 22 conventional
synchronous power plants with AVRs and turbine governors, 6 PSSs and
169 wind power plants.  Note that the secondary frequency control of
the AIITS is implemented manually and, thus, is slower than $200$ s,
hence no AGC is considered in the model.  Wind speeds are modeled as
OU processes.  The resulting set of DAEs for the AIITS includes 2278
state variables (666 of which are stochastic processes) and 14623
algebraic variables. The proposed approach was solved for all of the
state and algebraic variables but, for space limitation, we can show
below only a small selection of these variables.

\begin{figure}[b]
  \centering
  \includegraphics[width=0.9\linewidth]{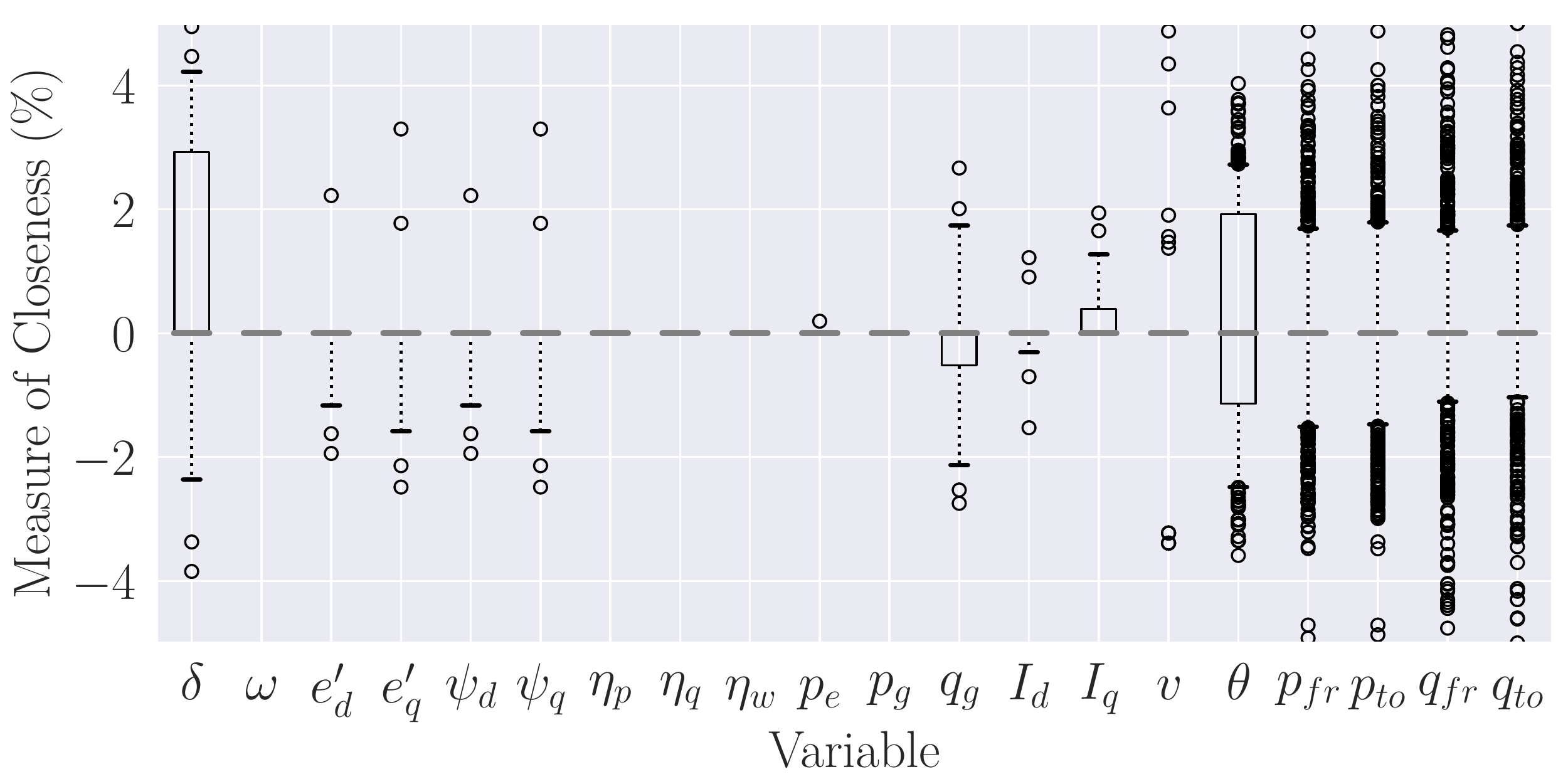}
  \caption{Box plot of the \MC{} $\epsilon_{\sigma}$ for the AIITS.}
  \label{fig:irish_box}
  \vspace{-2mm}
\end{figure}

The box plot of $\epsilon_{\sigma}$ for the AIITS is shown in
Fig.~\ref{fig:irish_box}.  The deviations observed in
$\epsilon_{\sigma}$ are of the same order as the $\epsilon_{\sigma}$
observed in IEEE 14-bus system.  This result shows that the LEM works
for larger systems with the same accuracy as it does for the smaller
systems.  It is important to note that the solution of
\eqref{eq:covy}, which is straightforward \textit{per se}, depends on
the solution of the Lyapunov equation in \eqref{eq:lyapunov}, which is
numerically more challenging, especially for large systems.  To
properly solve \eqref{eq:lyapunov}, the matrix $\bfb A_o$ has to be
well-conditioned to prevent numerical overflow but at the same time to
retain the accuracy of the results.

It is also relevant to note that the increase in $\epsilon_{\sigma}$
values does not necessarily need to be interpreted as an error of the
proposed approach.  Indeed, non-zero values in $\epsilon_{\sigma}$ can
be expected in general because $\epsilon_{\sigma}$ is a relative value
and none of the two methods (i.e., MC and LEM) is perfect in terms of
accuracy.
On the one hand, the MC method is a brute-force numerical technique.
Its accuracy depends on the number of realizations of the stochastic
processes and the time step of the time domain integration, as well as
the length of the simulated time.  On the other hand, the LEM is
analytical and exact, at least for linear or linearized systems.
Hence, the deviation between the results obtained with these two
techniques is due to (i) the numerical issues of the MC method and
(ii) the nonlinearity of the DAEs that model the power system
behavior.

LEM shows a clear advantage with respect to MC, at least for large
power system models.  That is, LEM is characterized by significantly
smaller computational times than MC and yields variances of algebraic
variables with higher accuracy.  In the case of the AIITS, the total
CPU time required by MC was $14763$ s, i.e., more than 4 hours,
whereas LEM took $12$ s.

\section{Conclusions}

This letter proposes a method to calculate the standard deviation of
the algebraic variables of power system modeled as \acp{sdae}.  The
proposed method is based on the solution of the Lyapunov equation and
a linearized method.  Simulation results show that the proposed
technique has a high accuracy for a wide range of standard deviation
of stochastic processes, and significantly reduces the computational
time compared to conventional Monte Carlo time domain simulations.
Moreover, the proposed method calculates the variance of \textit{all}
algebraic variables based on the knowledge of the variance of the
noise sources and the system model, such as load active and reactive
power and wind speeds.  These are typically known by system operators.
The proposed approach appears thus more practical than measuring
directly all the algebraic variables in the system.  Finally, since
the proposed approach is analytical, it makes possible a deeper
understanding of the phenomena under study.  For example, one can
easily and quickly run a sensitivity analysis by varying the elements
of $\bfb G_o$ or design a robust control that minimizes the effect of
the noise on certain algebraic variables. These analyses are not
straightforward or even possible using the MC method.

The limitations of this method are similar to those of any
linearization.  We expect that the accuracy of the approach reduces as
the time scale of the analysis and, hence, the deviation of the actual
system from the linearized model increases.  How the accuracy of this
approach varies as a function of the time-scale is indeed an open
question which we will tackle in future work.  In particular, we will
also focus on the evaluation of the impact of nonlinearities such as
saturations and controller hard limits on the variance of the
variables of stochastic power system models as well as on the design
of robust controller.

\end{document}